\documentclass[conference]{IEEEtran}
\IEEEoverridecommandlockouts
\usepackage{cite}
\usepackage{amsmath,amssymb,amsfonts}
\DeclareMathOperator{\E}{\mathbb{\E}}
\usepackage{algorithmic}
\usepackage{graphicx}
\usepackage{textcomp}

\usepackage{ulem}
\usepackage[utf8]{inputenc}
\usepackage{subcaption}
\usepackage{multirow}
\usepackage{array}
\usepackage{float}  
\usepackage{pgfplots}
\usepackage{subcaption}
\usepackage{soul}
\usepackage{balance}
\graphicspath {{Figures/}}

\def\BibTeX{{\rm B\kern-.05em{\sc i\kern-.025em b}\kern-.08em
    T\kern-.1667em\lower.7ex\hbox{E}\kern-.125emX}}

\begin{document}

\title{A Unified Deep Transfer Learning Model for Accurate IoT Localization in Diverse Environments}

\author{
  Abdullahi Isa Ahmed$^{\ddagger}$,
  Yaya Etiabi$^{\ddagger}$, Ali Waqar Azim$^{\ast}$, and
  El Mehdi Amhoud$^{\ddagger}$\\

$^{\ddagger}$College of Computing, Mohammed VI Polytechnic University (UM6P), Benguerir, Morocco \\

$^{\ast}$University of Engineering and Technology, Taxila, Pakistan \\

Emails: \{abdullahi.isaahmed, yaya.etiabi, elmehdi.amhoud\}@um6p.ma, aliwaqarazim@gmail.com}

\maketitle

\begin{abstract}

Internet of Things (IoT) is an ever-evolving technological paradigm that is reshaping industries and societies globally. Real-time data collection, analysis, and decision-making facilitated by localization solutions form the foundation for location-based services, enabling them to support critical functions within diverse IoT ecosystems. However, most existing works on localization focus on single environment, resulting in the development of multiple models to support multiple environments. In the context of smart cities, these raise costs and complexity due to the dynamicity of such environments. To address these challenges, this paper presents a unified indoor-outdoor localization solution that leverages transfer learning (TL) schemes to build a single deep learning model. The model accurately predicts the localization of IoT devices in diverse environments. The performance evaluation shows that by adopting an encoder-based TL scheme, we can improve the baseline model by about 17.18\% in indoor environments and 9.79\% in outdoor environments. 

\end{abstract}

\begin{IEEEkeywords}
Internet of Things (IoT), localization, deep learning, received signal strength indicator (RSSI), transfer learning.
\end{IEEEkeywords}

\section{INTRODUCTION}\label{intro}

Over the past few decades, there has been a significant increase, in the study of Internet of Things (IoT) localization solutions. Researchers in various industries are putting in efforts to enhance the accuracy and effectiveness of locating devices in different scenarios. Furthermore, it is expected that by 2030, the number of connected devices globally will be about 24.1 billion and will continue to increase \cite{b1}. In response to this massive growth, researchers have delved into a range of localization approaches incorporating several environmental settings to cater to the diverse requirements of IoT deployments.

In the context of IoT systems, the performance of IoT localization techniques and technologies depends on different environmental settings. These include indoor, outdoor, underground, urban, and suburban constraints\cite{b2}. However, most existing research focuses on solving problems in one specific environment, leading to the development of separate models for each environment. Indeed, numerous studies have explored indoor localization techniques utilizing technologies such as bluetooth low energy (BLE) beacons, ultra-wideband (UWB), radio frequency identification (RFID), Wi-Fi, visible light communication (VLC), to mention a few. In addition, a considerable amount of existing research has been published for indoor localization \cite{b3, b4, b5, b6}. These studies offer diverse benefits in various indoor settings, potentially enhancing efficiency and productivity in our daily lives. However, concerns about the generalizability of these systems arise due to limited testing across diverse settings.

On the other hand, outdoor localization has been investigated using global navigation satellite systems (GNSS) which include, among others, Galileo, Beidou, and global positioning system (GPS). However, the most prominent drawbacks of these technologies are the high power consumption and cost \cite{b7}. Thus, the widespread implementations of these technologies are hindered. As a consequence, there are high demands for new low-cost, low-power consumption, and long-range wireless technologies such as NarrowBand IoT (NB-IoT), long-term evolution machine type communication (LTE-M), Sigfox, and long-range wide area networks (LoRaWAN). As far as these demands are concerned, research works have shown the emerging long-range (LoRa) technology at the physical layer and its complementary LoRaWAN protocol at the upper layer brings forth notable benefits in terms of coverage, energy efficiency, and scalability \cite{b8}. Several authors have proposed frameworks to address the challenges of outdoor localization. The authors in \cite{b10, b11} have \mbox{discussed} the growing need for accurate localization solutions in outdoor location-based services. They propose deep learning (DL) framework to improve the existing works. Nevertheless, their proposed framework are indeed vulnerable to environmental changes. As a result, these methods are difficult to generalize across different scenarios.

Consequently, it is important to have a unified and seamless localization solution that covers both indoor and outdoor settings, as focusing solely on either indoor or outdoor \mbox{localization} can lead to limitations due to the dynamic nature of the environments. To address these challenges, recent attention has focused on developing two or more separate models specifically tailored for each environment. The authors in \cite{b12, b13} presented multiple deep neural network (DNN) systems for accurate indoor and outdoor localization. However, the use of multiple models for different environments can introduce an added layer of complexity in the overall systems.

In this paper, we are interested in developing an intelligent machine learning (ML) model for use in IoT applications such as smart parking, waste management, city asset tracking, emergency response etc. The aim is to accurately predict the location of these IoT sensors, which is crucial for the functionality of IoT applications. Existing localization solutions often target a single environment, which limits their effectiveness. To address this issue, we present a new localization-based DNN framework that unifies the localization process with a transfer learning (TL) approach. This framework eliminates the need for separate models for each environment, reducing development costs, complexity, and errors. The authors in \cite{b13bis} propose a transfer learning framework for indoor localization that reduces setup and increases scalability. However, the work only looks at indoor environments, which might not work outside due to the differences in signals. To the best of our knowledge and at the time of writing this paper, this is the first time a single framework has been developed to predict the location of IoT devices in various scenarios. Consequently, our main contributions can be summarized as follows.

\begin{itemize}
\item Two robust systems are proposed: 1) an encoder-based model that uses TL to transfer knowledge between different environments (indoor to outdoor and vice versa) and 2) a unified multilayer perceptron (U-MLP) model that effectively distinguishes the environment for seamless localization. 

\item We have extensively analyzed and pre-processed the datasets used in this paper to gain insights, which has greatly aided in reducing localization errors.

\item We have examined the TL capability of our proposed model in two distinct environments.

\item Furthermore, our findings indicate that the proposed encoder-based TL approach enhanced the baseline model by approximately 17.18\% in indoor environments and 9.79\% in outdoor environments. This resulted in mean distance errors (MDEs) of 6.65m and 361.21m, respectively. Additionally, we propose a  U-MLP model that achieves MDEs of 9.61m and 341.94m in indoor and outdoor environments, respectively.
\end{itemize}

The rest of the paper is organized as follows. Section~\ref{proposed_system} presents the proposed architecture of the system model, followed by the performance evaluation and detailed analysis of the obtained result in Section~\ref {performance_evaluation}. Finally, the paper provides a comprehensive conclusion and set forth some perspectives in Section~\ref{conclusion}.

\section{PROPOSED SYSTEM ARCHITECTURE}\label{proposed_system}

In this work, the principal objective is to propose an optimal DNN framework that will predict the location $\mathit{L}$ of a connected IoT device in diverse environments given a set of input parameters, i.e., the received signal strength indicator (RSSI) values. To achieve this, we consider the set RSSI signals from both \mbox{Wi-Fi} and LoRa networks in indoor and outdoor environments, respectively. The indoor environment consists of a set of $\mathcal{W}$ access points (APs) for Wi-Fi signals, while the outdoor environment utilizes a set of $\mathcal{G}$ LoRa gateways (GWs) to establish localization for IoT devices.

\subsection{Encoder Network}
The architecture presented in Fig.~\ref{fig1a} uses an \mbox{encoder-based} TL framework. The encoder, which is a type of artificial neural network (ANN) \mbox{framework}, aims to learn an efficient representation of the input data by transforming it into a latent space. This latent representation, denoted as $\mathcal{Z}$, is defined in a Euclidean space, such that $\mathcal{Z} \in \mathbb{R}^{n}$. More precisely, the encoder $\xi_{\phi}: \mathcal{X} \to \mathcal{Z}$ is parameterized by $\phi$, where $\phi$ represents a neural network (NN) approximator. In our model, the encoder, represented by $\xi_{\phi}$, acts as a bottleneck, mapping an input space $\mathcal{X} \in \mathbb{R}^{\bar{\mathcal{W}}} \cup \mathbb{R}^{\bar{\mathcal{G}}}$ to a latent space $\mathcal{Z}$. Here, the input space, denoted by $\mathcal{X}$, can be represented by either a real vector of dimension $\bar{\mathcal{W}}$ or $\bar{\mathcal{G}}$ which correspond to the preprocessed RSSI measurements from the Wi-Fi and LoRa networks, respectively. This transformation is expressed as:

\begin{equation}
z = \xi_{\phi}(x),
\label{eqt:encoder}
\end{equation}
where $z$ represents the latent representation. Specifically, our encoder is implemented as an $n$-layer ANN where the prediction for each hidden layer $h$ is given by:

\begin{equation}
x_{h+1} = \sigma_h(W_h^T x_h + b_h), h= {0, 1, \dots, n-2}
\label{eqt_hidden}
\end{equation}
where the $\sigma_h(\cdot)$ is an activation function applied at layer $h$. $W_h$ is the weight matrix for layer $h$. The bias vector for layer $h$ is represented by $b_h$. Note that in Eq. (\ref{eqt_hidden}), $x_0$ is the initial input to the encoder and correspond to the input data $x$ while $x_{n-1}$ represents the output of the final hidden layer, which is denoted as the latent representation $z$.

\subsection{Transfer Learning in Diverse Environment}\label{transfer_learning}

Unlike conventional machine learning (ML) techniques that train independent models for each task, TL leverages knowledge from a source task $S$ to enhance learning in a target task $T$, even when they exhibit distinct data patterns or feature sets. The input and output spaces for $S$ are represented by $\mathcal{X}_S$ and $\mathcal{Y}_S$, respectively, while the corresponding spaces for $T$ are denoted by $\mathcal{X}_T$ and $\mathcal{Y}_T$. Specifically, this task aims to find a function $f(\cdot)$, that minimizes the expected loss on the source environment that is given by:
\begin{equation}
L(\theta^{S}) = \mathbb{E}_{(x_S, y_S) \sim \mathcal{P}_S} [\ell(f(x_S; \theta^{S}), y_S)],
\label{eq:min_expected_loss}
\end{equation}
where $\theta^{S}$ is the source model parameters. $x_S$ denotes an input data point from the source task distribution. $y_S$ represents the source value associated with $x_S$. $\ell$ is the loss function that measures the model performance. $\mathcal{P}_S$ signifies the source task data distribution, $\mathbb{E}_{\left\{ \cdot  \right\}}$ denotes the expectation operation, and $L(\theta^{S})$ is the expected loss on the source domain.
 
\begin{figure}[t]
    \centering
    \begin{subfigure}{\linewidth}
        \centering
        \includegraphics[width=\linewidth]{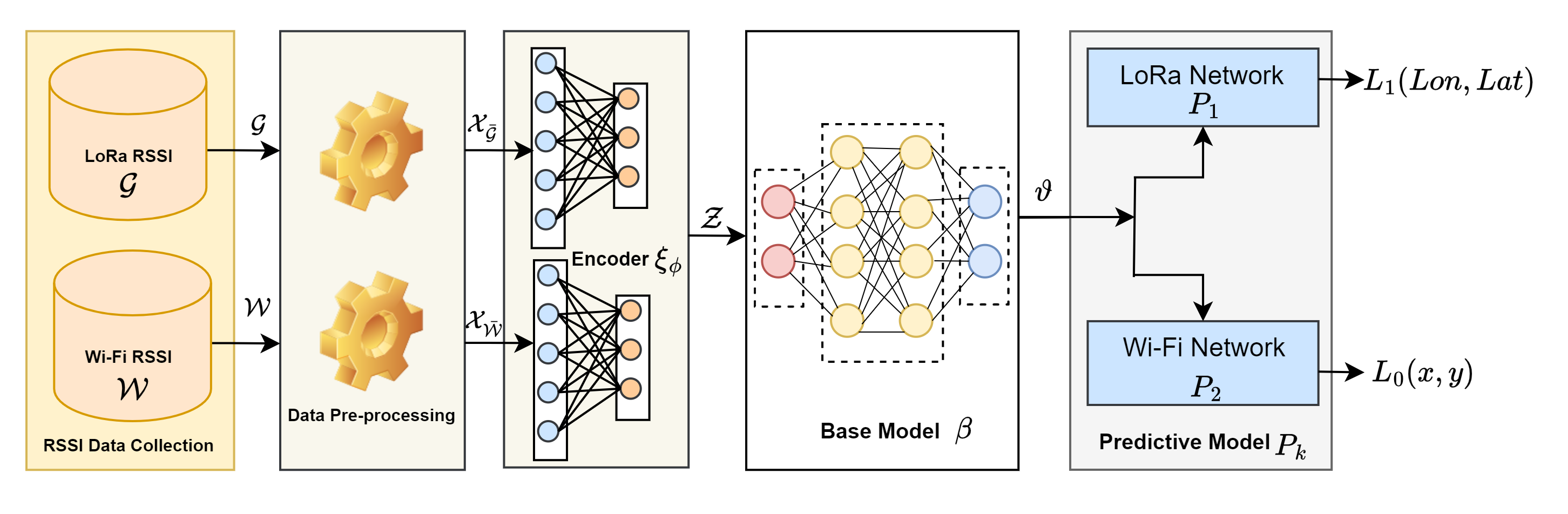}
        \caption{Proposed Encoder-based system.}
        \label{fig1a}
    \end{subfigure}
    \begin{subfigure}{\linewidth}
        \centering
        \includegraphics[width=\linewidth]{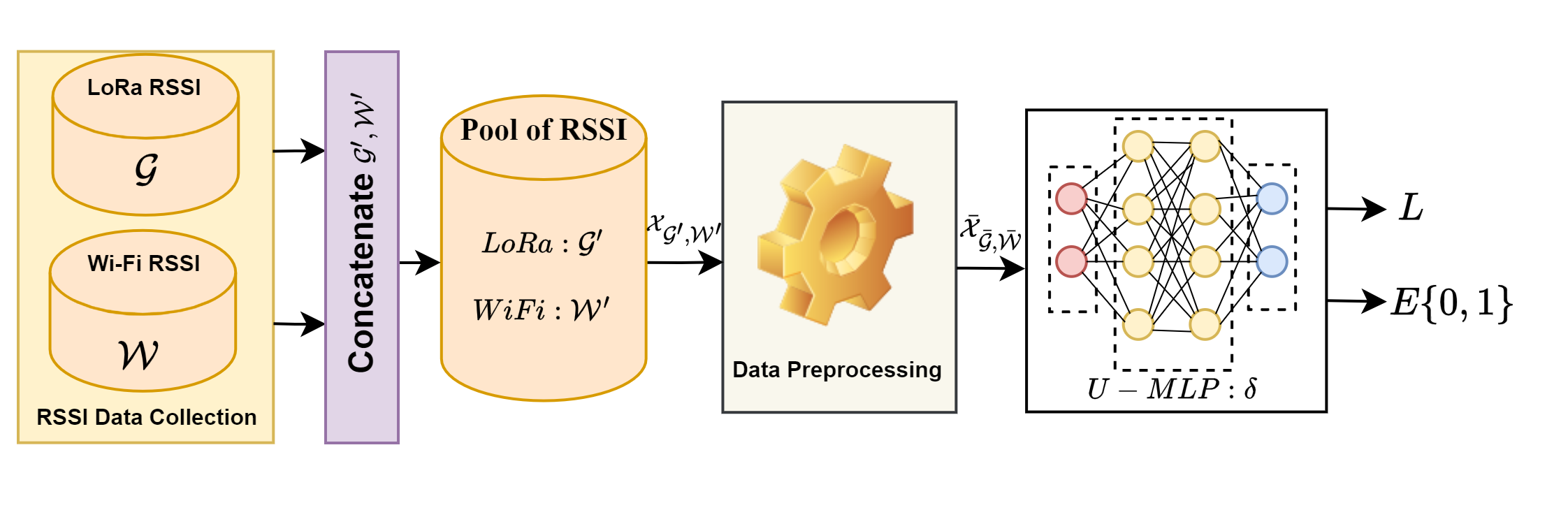}
        \caption{Proposed U-MLP system.}
        \label{fig1b}
    \end{subfigure}
    \caption{Multi-Environment model-based indoor and outdoor localization system.}
    \label{fig1}
\end{figure}

As shown in Figure~\ref{fig1a}, our model utilizes an encoder $\xi_{\phi}$ framework to transform the input data ($x_S$ or $x_T$) into a fixed latent representation. This is necessary due to the varying input shapes resulting from differences in the number of GWs or APs. Thus, the encoder ensures a common latent representation for our diverse environments. Furthermore, the predictive model denoted by $P_k$ where $k\in \{ {P_{1}, P_{2}} \}$ corresponded to LoRa and WiFi networks, respectively, as shown in our proposed system (Fig.~\ref{fig1a}). $P_k$ takes as input $\vartheta$ (the output of our base model $\beta$) and outputs the fine-grained 2D location of either indoor or outdoor positioning $L_0$ and $L_1$ that corresponds to Wi-Fi and LoRa networks, respectively. These networks both utilize $\hat{\sigma}_{w}(\cdot)$ and $\hat{\sigma}_{l}(\cdot)$ linear activation functions. 

Therefore, the model $\theta^{S}$ in our source environment was initialized by $\theta^{S}_{0} = (\xi^{S}_{0}, \beta^{S}_{0}, P^{S}_{k0})$, where the initialized encoder framework is represented by $\xi^{S}_{0}$, $\beta^{S}_{0}$ is the base model, and $P^{S}_{k0}$ is the predictive model. After training the $\theta^{S}_{0}$ model by minimizing the loss in Eq.(\ref{eq:min_expected_loss}), we ended up with the optimal parameters \mbox{$\theta^{S}_{*} = \underset{\theta^{S}}{argmax}L(\theta^{S})$}. Note that $\theta^{S}$ is trained using stochastic gradient descent (SGD). For clarification, let recall that \mbox{$\theta^{S}_{*} = (\xi^{S}_{*}, \beta^{S}_{*}, P^{S}_{k*})$}. The key idea is to use the pre-trained parameters of the base model $\beta^{S}_{*}$ as a starting point for training our model $\theta^{T}$ in the target environment. In the target environment, the model is initialized by \mbox{$\theta^{T}_{0} = (\xi^{T}_{0}, \beta^{S}_{*}, P^{T}_{k0})$} and we aim to get an approximation to the optimal parameters $\theta^{T}_{*}$ by iteratively running $\mathcal{N}$ steps of SGD using the following equation:

\begin{equation}
\theta^{T}_{i+1} = \theta^{T}_{i} - \alpha \nabla L(\theta^{T}_{i}), i={1, \dots, \mathcal{N}-1}
\label{eq:GD}
\end{equation}
where $\alpha$ is the learning rate of the gradient descent, $\nabla L(\cdot)$ represents the gradient of the loss function $L$ evaluated at $\theta^T_i$ and $i$ is the iteration counter from 1 to $\mathcal{N}-1$.

Additionally, we design a unified multilayer perceptron (U-MLP) model $\delta$, as shown in Fig.~\ref{fig1b}. This model $\delta$ uses fully connected dense layers for training and it incorporates features from diverse environments. The input layer of the system consists of a collection of RSSI data from both LoRa $\mathcal{G}$ and Wi-Fi $\mathcal{W}$. We sample an equal number of data points from $\mathcal{G}$ and $\mathcal{W}$ and we randomly concatenated them to have a balance RSSI values for these two diverse data. Similarly, this single system is configured to function as a multitask network. This is because it simultaneously performs a regression  and classification task by predicting the environment $\mathit{E}$ and estimating the coordinate $\mathit{L}$. The predicted environment $\mathit{E}$, where $E \in \{0, 1\}$, i.e., 0s for indoors and 1s for outdoors, while the $\mathcal{L}$ coordinates contain either the normalized coordinate of $Lon, Lat$ for outdoor or reference point of $x, y$ for indoor environment. 

\subsection{RSSI Data Collection}
The datasets used in this paper were obtained from \cite{b14} and \cite{b15}, for the indoor and outdoor settings, respectively. The two datasets constituted RSSI measurements from different GWs and APs. The total area in \cite{b14} is approximately 108,703 square meters, including three buildings, each having four or five floors. There are 933 distinct reference points in the database. In all, there were 21,049 sampled points in total. More than 20 users used 25 different models of mobile devices to collect data, with some users using multiple device models utilizing WiFi technology. Similarly, in \cite{b15}, the LoRaWAN dataset was collected for outdoor fingerprinting in Antwerp, Belgium. This dataset contains 130,429 messages, with each message including RSSI signal strength data from 72 GWs, GPS-based ground truth locations in terms of $\mathit{Lon}$ and $\mathit{Lat}$, and information on the LoRa spreading factors (SF), which plays a role in signal range and it varies from 7 to 12.

\subsection{Data Preprocessing}\label{data_pre}

A closer inspection of the datasets was carried out separately and it appears that not all GWs and APs are within the range during measurement. Thus, the missing RSSI signals were set to 100dBm and -200dBm at the time of collection to indicate that no signal was received for indoor and outdoor, respectively. This indicates a significant amount of outliers in the data. To handle these missing RSSI values and outliers in the indoor data, we adopted similar concepts as in \cite{b16}, by setting a threshold of 98\% for those APs (features) that contain the missing values and dropping out any APs above this threshold measurement. At this point, we were left with only 249 APs as compared to the initial 520 APs. Furthermore, the outdoor dataset was analyzed, and due to the limitation of the dataset, we decided to adopt the entire 72 GWs corresponding to the features for our model deployment. We also handled the missing RSSI Values by replacing them with a -128dBm to avoid outliers and we applied a custom MinMax normalization technique as defined below.

\begin{equation}
    RSSI_{i}^n  = \left\{\begin{matrix}
    \; &0, \text{ if } RSSI_{i} \text{ is not valid}\\ \\
    \;&a+\left ( \frac{RSSI_{i} - RSSI_{min}}{RSSI_{max} - RSSI_{min}} \right )\left ( b-a \right ), \text{ Otherwise}
\end{matrix}\right..
\label{eqt1}
\end{equation}

Eq.(\ref{eqt1}) produces from the original $RSSI_{i}$ a positive and normalized representation $RSSI_{i}^n $ ranging from $a$ the lower bound of the normalized $RSSI$ to $b$ the upper bound, which are set empirically to improve the performance of the model. Similarly, we use the MinMax scaling process to ensure the latitude and longitude values fall within the range of 0 and 1.

\subsection{System Evaluation Metrics} 
To evaluate our system performance, we have applied two different metrics to evaluate the robustness of our models, depending on the nature of the dataset. For example, the indoor or rather the Wi-Fi model uses the MDE to assess the accuracy of predicted localization errors as shown in Eq.(\ref{eqt2a}). While the Haversine formula was used to evaluate our outdoor framework as defined in Eq.(\ref{eqt3}). The MDE metric is used in localization tasks to calculate the average Euclidean distance error between the predicted location and the actual location. It assesses how accurately our model predicts the spatial positions. This can be expressed as follows:

\begin{equation}
\text{MDE} = \frac{1}{N} \sum_{i=1}^{N} \sqrt{(x_i - \hat{x}_i)^2 + (y_i - \hat{y}_i)^2},
\label{eqt2a}
\end{equation}
where $N$ represents the total number of data points and the summation is carried out over each data point indexed by $i$. The variables $x_i$ and $y_i$ represent the actual 2-D coordinates for the $i$-th data point, respectively, while $\hat{x}_i$ and $\hat{y}_i$ represent the predicted 2-D coordinates for the $i$-th data point, respectively. 

In addition, the Haversine formula given in Eq.(\ref{eqt3}) was utilized to measure distance errors in the outdoor dataset. The geodesic distance between two points on a curved surface such as the Earth was calculated. Thus, this was chosen due to its suitability for the outdoor scenario. 

{\begin{equation}
d = 2R \arcsin\Bigg(\sqrt{\gamma + cos\phi_{1}cos\phi_{2}sin^{2}(\frac{\lambda_{2}-\lambda_{1}}{2})}\Bigg),
\label{eqt3}
\end{equation} 
}
where $d$ is the distance between the two points in kilometers, $R$ is the radius of the earth (i.e., 6371.01 kilometers), and $\lambda_1$ and $\lambda_2$ are the longitudes of the two points in degrees. Similarly,  $\gamma = sin^{2}\big(\frac{(\phi_{2}-\phi_{1})}{2}\big)$, where $\phi_1$ and $\phi_2$ are the latitudes of the two points in degrees.

\section{PERFORMANCE EVALUATION AND DISCUSSION}\label{performance_evaluation}

\subsection{Simulation Setup and Parameters}

In order to study the performance of the proposed approach, an experiment was conducted using open-source ML and deep learning (DL) frameworks based on Python. Furthermore, Table~\ref{Table_1.0} shows the global model configuration of our proposed system architecture. The simulation was carried out using the Adam optimizer with a learning rate of $5 \times 10^{-4}$. In addition, a rectified linear unit (ReLU) activation function was applied to all intermediate layers. For model training, a batch size of 256 and 512 were adopted for Wi-Fi and LoRa data, respectively. Batch normalization was employed after each layer to enhance training stability and prevent overfitting, ensuring normalized inputs with a mean of zero and a variance of one. The loss function selected for these models is the mean square error (MSE), with root mean square error (RMSE) employed as the evaluation metric.

\begin{table}
\Huge
\centering
\caption{Simulation parameters of the system model.}
\label{Table_1.0}
\resizebox{\columnwidth}{!}{%
\begin{tabular}{clc|clc}
\hline
\multicolumn{3}{c|}{\textbf{(a) Encoder $\xi$}} & \multicolumn{3}{c}{\textbf{(b) Base Model $\beta$}} \\ \cline{1-3} \cline{4-6} 
\textbf{Symbol} & \textbf{Description} & \textbf{Value} & \textbf{Symbol} & \textbf{Description} & \textbf{Value} \\ \cline{1-3} \cline{4-6} 
$h_{\xi}$ & Hidden layers & $\{64,256,512\}$ & $h_{\beta}$ & Hidden layers & $\{32,64,128,512\}$\\
$D_{\xi}$ & Dropout & $0.1$, $0.2$ & $D_{\beta}$ & Dropout & $0.05$, $0.1$, $0.2$ \\
$BN_{\xi}$ & Batch Normalization & Yes & $BN_{\beta}$ & Batch Normalization & Yes \\
$\sigma_{\xi}(\cdot)$ & Activation functions & ReLU & $\sigma_{\beta}(\cdot)$ & Activation functions & ReLU \\
$\mathcal{Z}_{\xi}$ & Latent dimension & $128$ & $\mathcal{C}_{\beta}$ & Intermediatory dimension & $64$ \\ \cline{1-3} \cline{4-6} 
\multicolumn{3}{c|}{\textbf{(c) Wi-Fi Networks $P_{2}$}} & \multicolumn{3}{c}{\textbf{(d) LoRa Networks $P_1$}} \\ \cline{1-3} \cline{4-6} 
$h_{w}$ & Hidden layers & $\{32,128,256,512\}$ & $h_{l}$ & Hidden layers & $\{32,128,256,512\}$ \\
$D_{w}$ & Dropout & $0.015$, $0.1$ & $D_{l}$ & Dropout & $0.015$, $0.1$ \\
$BN_{w}$ & Batch Normalization & Yes & $BN_{l}$ & Batch Normalization & Yes \\
$\hat{\sigma}_{w}(\cdot)$ & Output activation function & linear & $\hat{\sigma}_{l}(\cdot)$ & Output activation function & linear \\
$\mathcal{Z}_{w}$ & Latent dimension & $150$ & $\mathcal{Z}_{l}$ & Latent dimension & $150$ \\
Optimizer & Model optimizer & Adam & Optimizer & Model optimizer & Adam \\
$\alpha_{w}$ & Learning rate & $5 \times 10^{-4}$ & $\alpha_{l}$ & Learning rate & $5 \times 10^{-4}$ \\
$e_{w1}, e_{w2}$ & Exponential decay rates & $0.1$, $0.99$ & $e_{l1}, e_{l2}$ & Exponential decay rates & $0.1$, $0.99$ \\
$B_{w}$ & Batch size & $256$ & $B_{l}$ & Batch size & $512$ \\ \hline
\end{tabular}%
}
\end{table}

\subsection{Data Exploration and Characterization} 

\begin{figure}[t]
    \centering
        \begin{subfigure}[t]{0.40\textwidth}
        \centering
        \includegraphics[width=\textwidth]{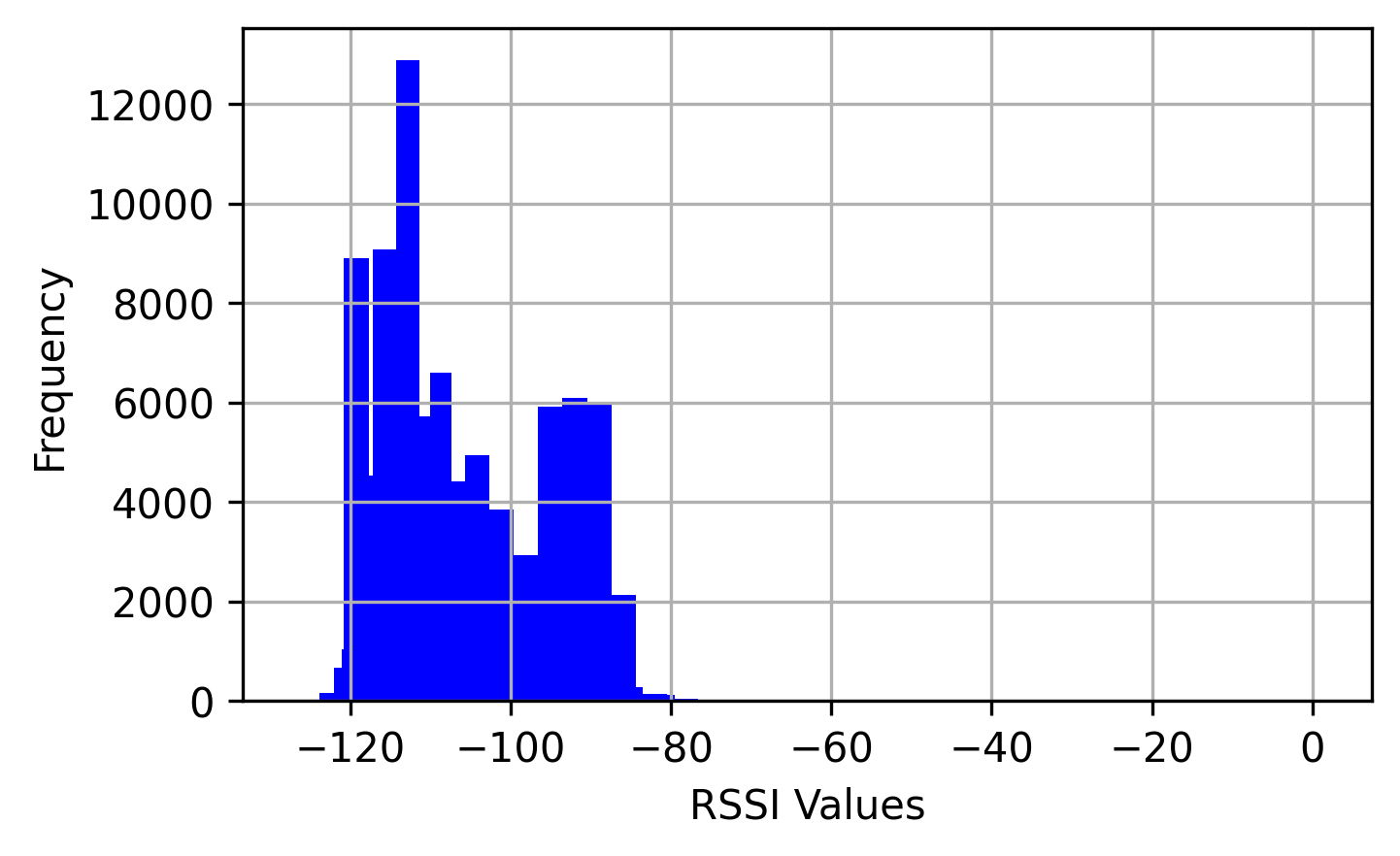}
        \caption{RSSI values for outdoor measurements (in dBm)}
        \label{fig2a}
    \end{subfigure}%
    \hfill%
    \begin{subfigure}[t]{0.40\textwidth}
        \centering
        \includegraphics[width=\textwidth]{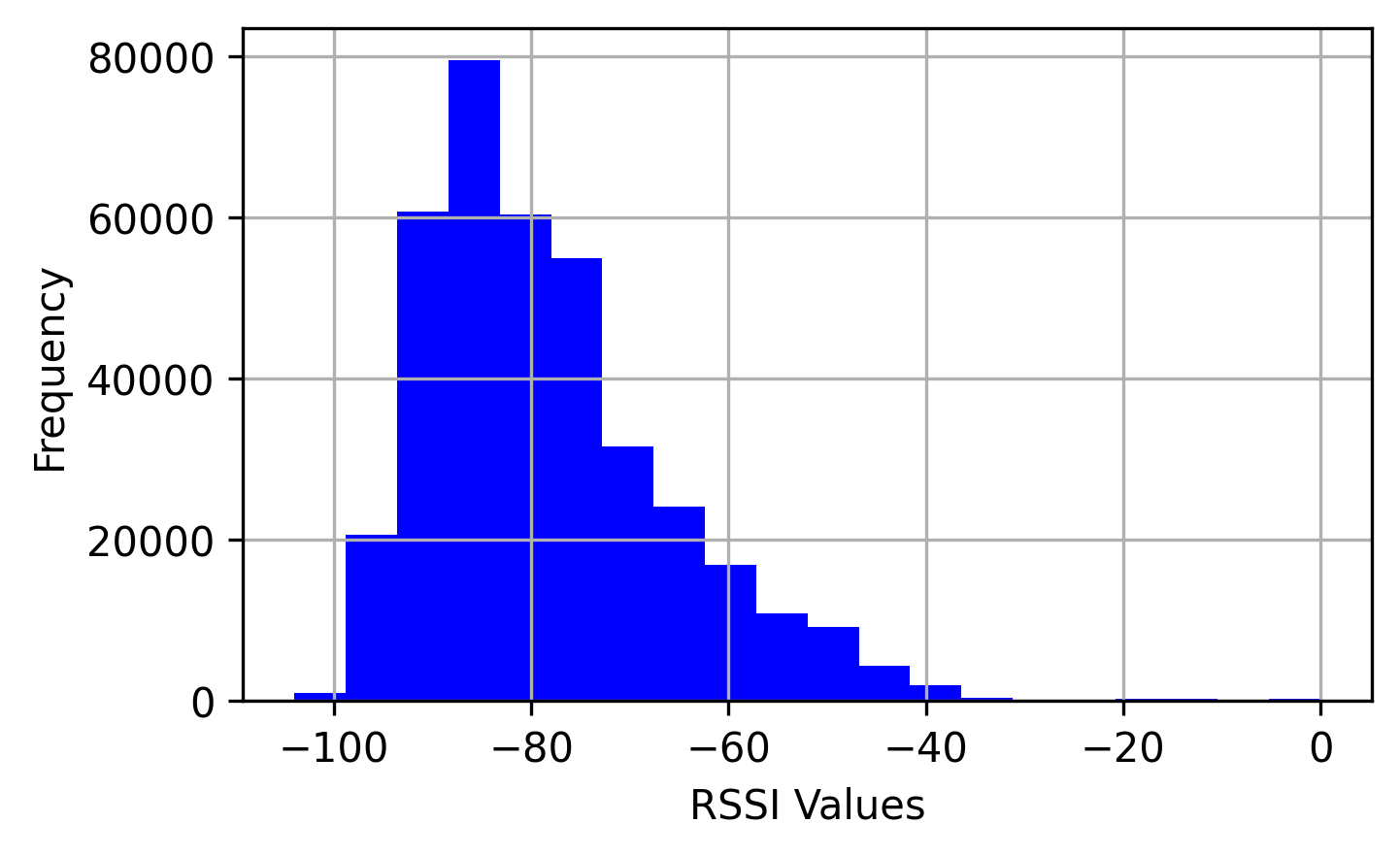}
        \caption{RSSI values for indoor measurements (in dBm)}
        \label{fig2b}
    \end{subfigure}
    \caption{Frequency of RSSI Values of Indoor-Outdoor database.}
    \label{fig2}
\end{figure}

Figure~\ref{fig2} shows the frequency distribution of the RSSI dataset across different intervals. Specifically, Fig.~\ref{fig2a} displays the outdoor RSSI with signal strengths ranging from -127 dBm to -60 dBm. Values of -200 dBm were added for base stations (BSs) not receiving signals, marked as out of range, and hence excluded from the histogram. The observed figure indicates a significant signal rise in the distribution from -120 to -100 dBm. The reason for this is the transmission over long distances in the LoRa setting and the high sensitivity of LoRa devices. In contrast, Fig.~\ref{fig2b} shows the distribution of RSSI values for indoors across the dataset. It is worth noting that this range of signal strengths varies from -104 dBm to 0 dBm. Additionally, a signal of 100 dBm was assigned to indicate that no signal was received at a certain wireless access point. However, only about 1.7\% of the entire recorded RSSI levels present in the database fall within the specific range of [-40 dBm to 0 dBm].


\subsection{Evaluation of the Encoder-based TL framework}

\begin{figure}[t]
    \centering
        \begin{subfigure}[t]{0.40\textwidth}
        \centering
        \includegraphics[width=\textwidth]{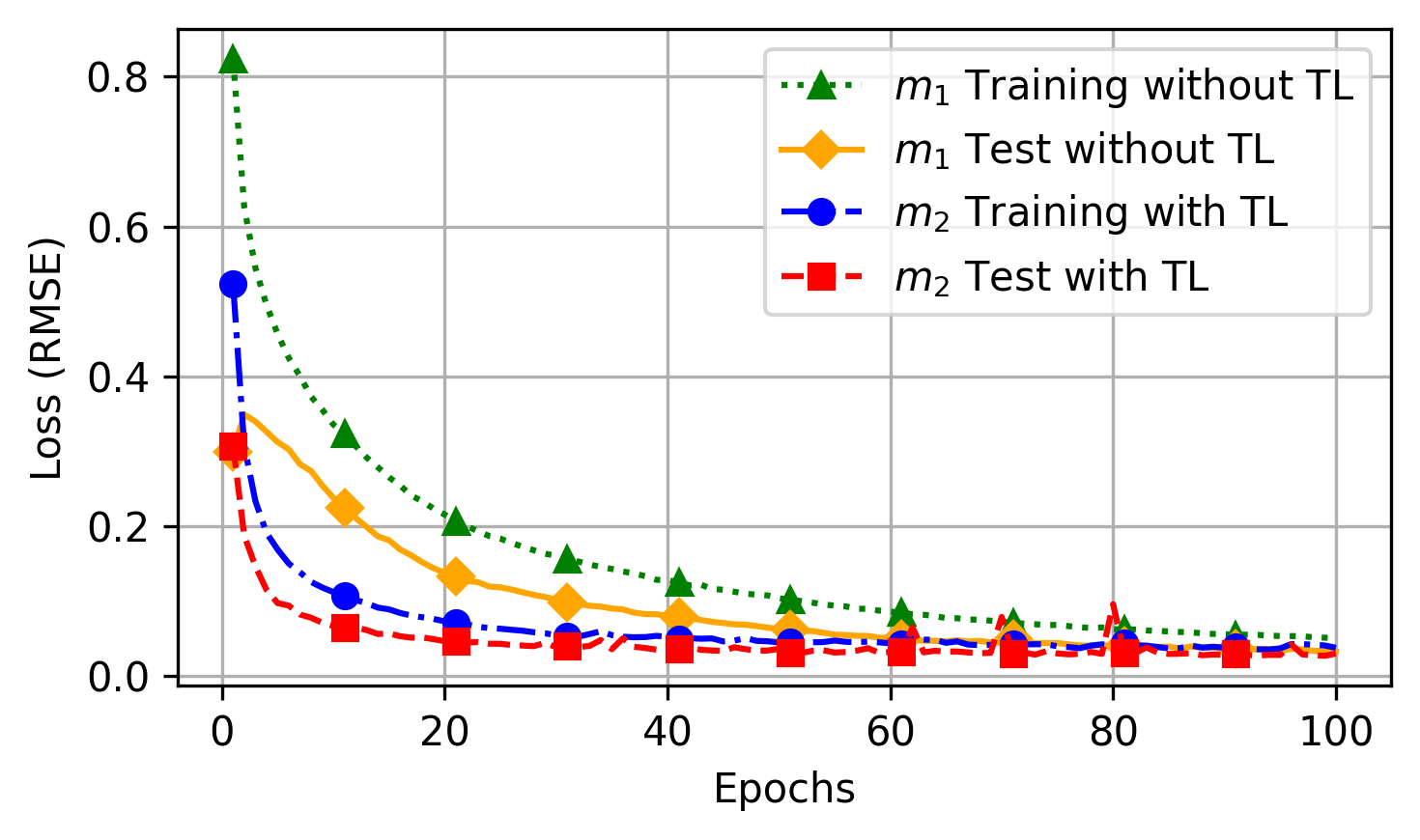}
        \caption{Indoor Environment.}
        \label{fig3a}
    \end{subfigure}%
    \hfill%
    \begin{subfigure}[t]{0.40\textwidth}
        \centering
        \includegraphics[width=\textwidth]{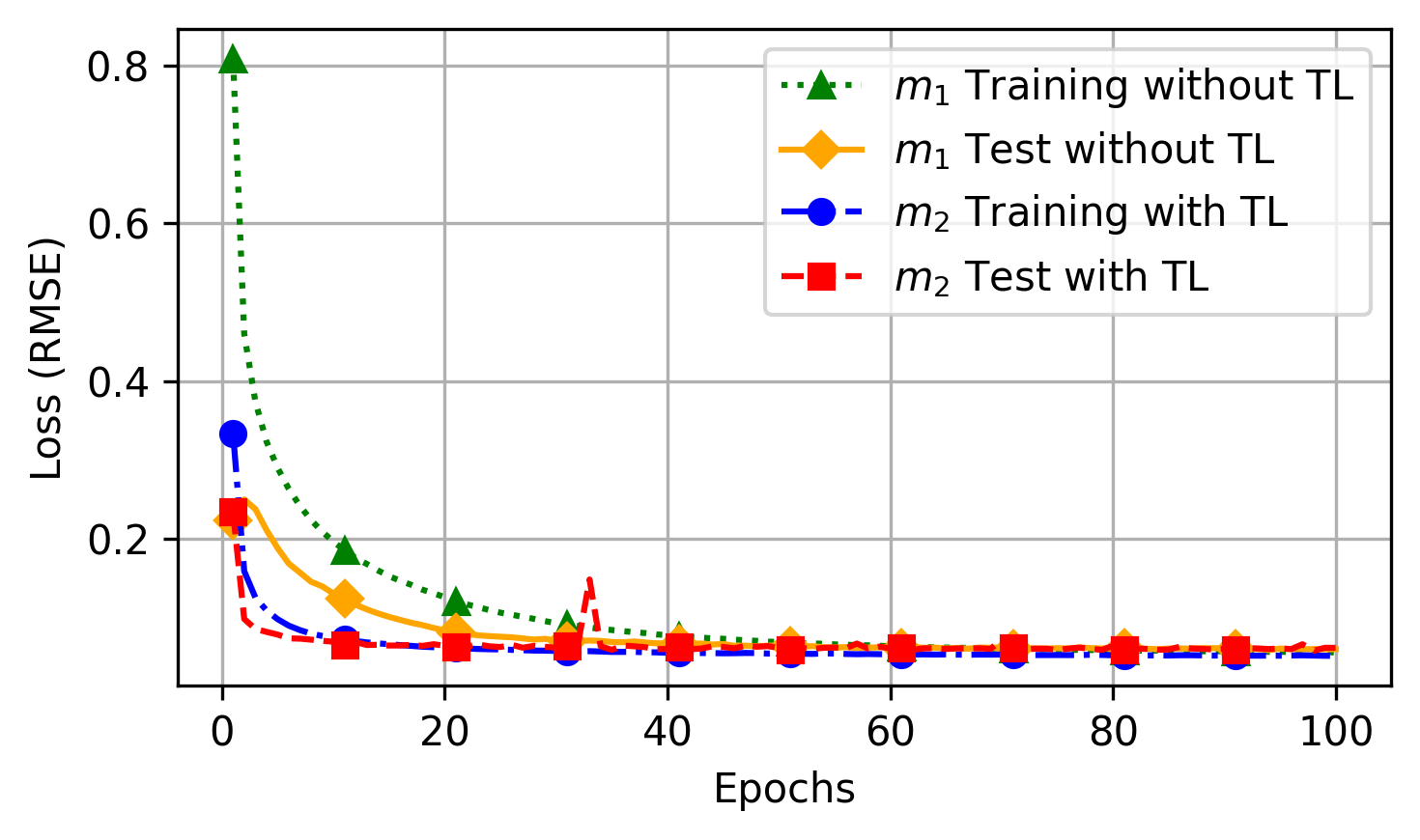 }
        \caption{Outdoor Environment.}
        \label{fig3b}
    \end{subfigure}
    \caption{Performance of Encoder-based TL Models in Indoor-Outdoor environment.}
    \label{fig3}
\end{figure}

In this part of the experiments, we evaluate the efficiency and scalability of the proposed encoder-based TL framework. As described in section~\ref{transfer_learning}, we defined the models in the source and target environment to be $\theta^{S}$ and $\theta^{T}$, respectively. In this section, our source and target environment are denoted by ${m}$ and ${m_2}$, respectively. First, we develop an ${m}$ model that was trained on $RSSI$ data from an outdoor environment. In addition, we developed two other models, $m_1$ and $m_2$. These two models were trained on the indoor environment data. The $m_1$ model was developed from scratch (i.e., without TL), and in $m_2$, rather than developing it from scratch, we utilize the learnable weight of the base model $\beta^{S}_{*}$ in ${m}$ to initialize $m_2$ (i.e., with TL). Finally, we validate the efficiency of our proposed encoder-based TL by comparing the two models $m_1$ and $m_2$. We plot the training and test performance of our models  based on the loss (RMSE) versus the number of iterations. As shown in Fig.~\ref{fig3a}, we notice that $\mathit{m}_2$ model converge faster. Specifically, it only takes the model with TL $\mathit{m}_2$ about 20 iterations to converge while about 50 iteration for $\mathit{m}_1$ to converge. Moreover, in Fig.~\ref{fig3b}, ${m}$ was trained in an indoor data while $m_1$ and $m_2$ were trained on the outdoor environment data. The model with TL converges faster with fewer epochs than the model without TL. 

In addition, our proposed encoder-based TL scheme demonstrates the scalability of our model. Unlike traditional ML approaches that require large and unique datasets for each new environment, our scheme leverages knowledge from a single pre-trained model, $m$. This results in a reduction of the data required to train for new environments, such as $m_1$ and $m_2$, leading to faster model training, and improved adaptability.

\subsection{Evaluation of the U-MLP framework}

\begin{figure}[h]
    \centering
    \begin{minipage}{0.4\textwidth}
        \centering
        \includegraphics[width=\linewidth]{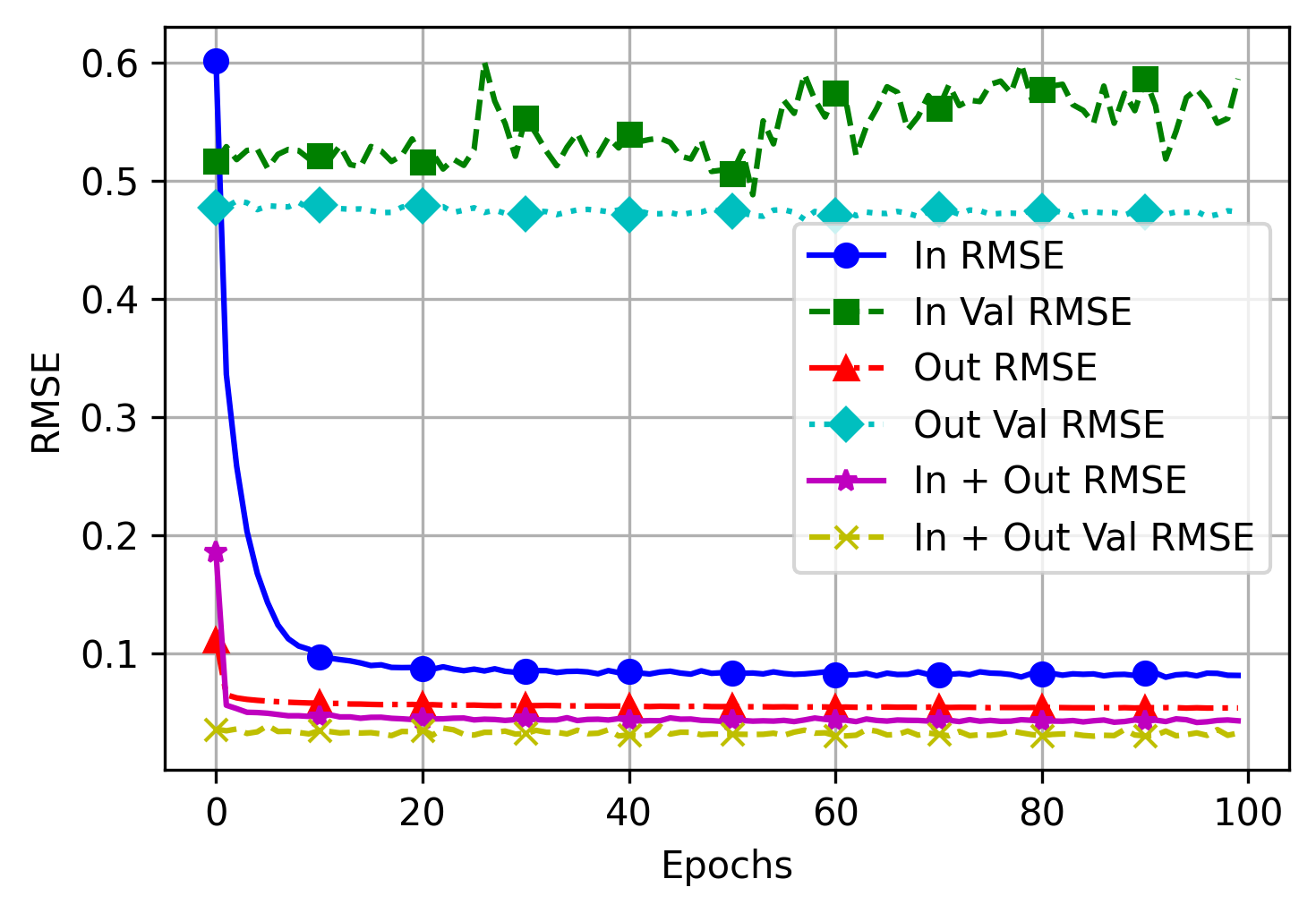}
        \caption{RMSE of MLP Model on Indoor, Outdoor, and Combined Datasets.}
        \label{fig4}
    \end{minipage}
    \hfill
    \begin{minipage}{0.4\textwidth}
        \centering
        \includegraphics[width=\linewidth]{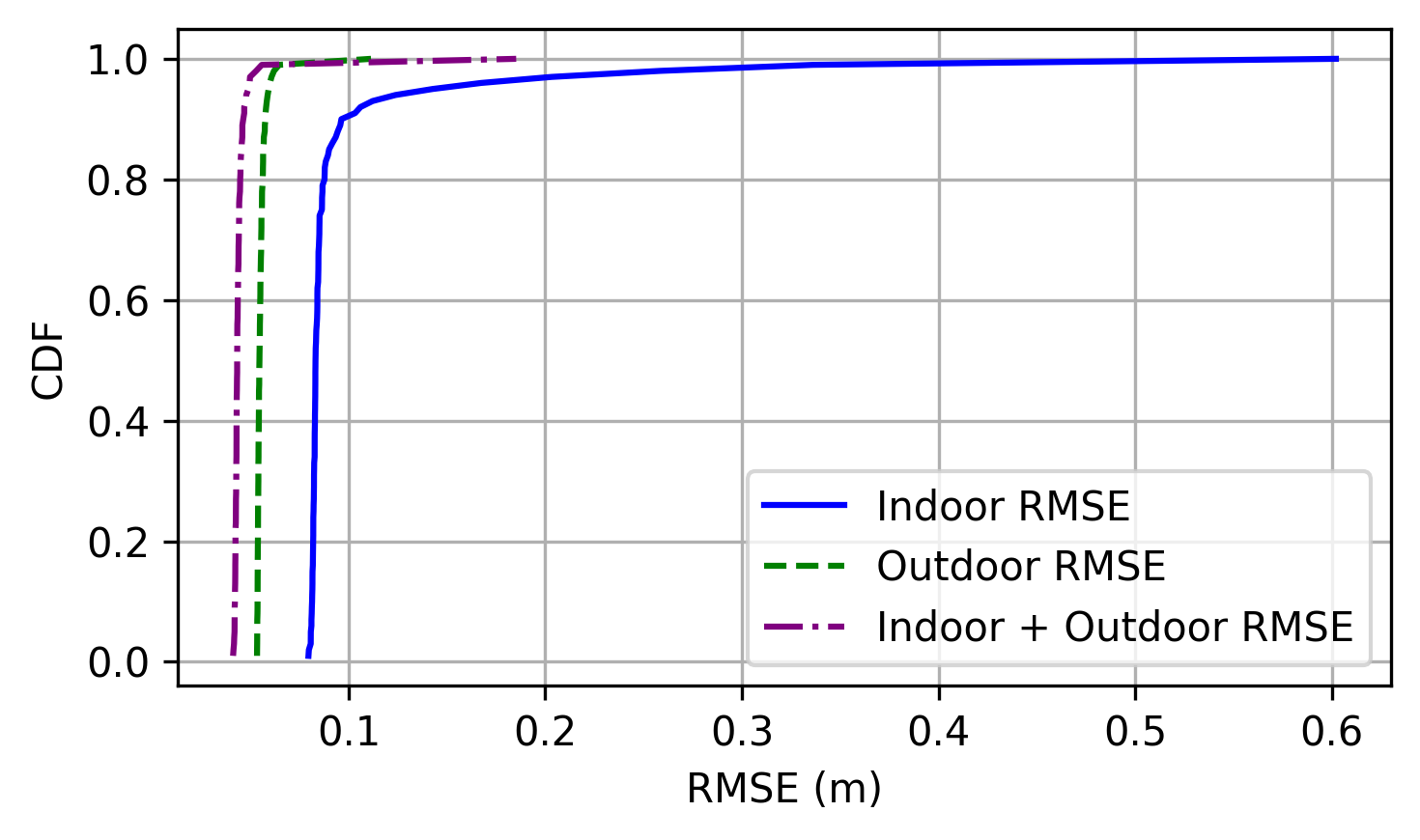}
        \caption{CDFs of RMSE for Indoor, Outdoor and Combined Datasets.}
        \label{fig5}
    \end{minipage}
\end{figure}

In this set of experiments, we train the previously deﬁned U-MLP model $\delta$ in Fig.~\ref{fig1b}. To achieve this implementation, we adopted similar conﬁgurations as shown in Table~\ref{Table_1.0}(b). The input RSSI data was based on a concatenated and equal samples obtained from both environments. After training, the framework further handles the separation of the predicted data. It utilizes the predicted probabilities from $\mathit{E}$ classes to differentiate between these diverse environments. The classification task achieves a prediction accuracy of 100\% which aid in a seamless localization. To verify we have the optimal model, we examine two separate environments (i.e. indoor and outdoor) and compare them with our pool of RSSI data (In + Out). To validate our methodology, we separately train indoor and outdoor models and validate them on the combined dataset. The learning curve of this system is depicted in Fig.~\ref{fig4} where it can be seen that the RMSE for indoor and outdoor environments has extremely low training errors and a very high validation error. Hence, these two models' performances are poor, and it signifies overfitting. It should be noted that the validation RMSE learning curve of the In model (sky blue line) remains relatively constant with little variation. This could be due to the larger size of our U-MLP model. In contrast, our U-MLP model \mbox{(In + Out)} has extremely low training and validation errors. Thus, achieving higher performance. 

Furthermore, we plot the cumulative distribution function (CDF) of the localization errors and compare it with three models, and the result is shown in Fig.~\ref{fig5}. For a fair comparison of the three models (i.e., In, Out and In+Out), the same configuration settings were used. Similarly, the x-axis which represents the localization RMSE error is based on the normalized data. This is due to the differential scaling in the errors of indoor environment which ranges within tenths of meters while the outdoor environment ranges within hundredths. As observed from the figure, the version with the combined (In + Out) framework further demonstrates that U-MLP framework has effectively achieves lower localization error compared with its counterpart.

Table~\ref{Table_2.0} presents a benchmark analysis of our two proposed framework and the models developed in \cite{b14} and \cite{b15} were employed as a benchmark for comparison. Furthermore, the encoder-based TL model we proposed achieves an MDE of 6.65m and 361.21m for indoor and outdoor environments, respectively. This represents an improvement of about 17.18\% in an indoor setting and 9.79\% in the outdoor setting compared to the baseline model. Similarly, our proposed U-MLP achieves 9.61m and 341.94m in the indoor and outdoor environment, respectively. 

It is important to note that our proposed approaches are further compared with the state-of-the-art models, as shown in Table~\ref{Table_2.0}. All of the compared models are based on a single environment approach (indoor or outdoor), whereas our proposed systems cover both indoor and outdoor environments. Furthermore, the proposed encoder model demonstrates superior performance in the indoor environment while maintaining competitive accuracy outdoors. Similarly, the proposed U-MLP model outperforms in outdoor settings while maintaining competitive accuracy indoors.

\begin{table}[h]
\caption{Comparison of the positioning performance of our proposed models with state-of-the-art.}
\label{Table_2.0}
\resizebox{\columnwidth}{!}{%
\begin{tabular}{lc|l|lc}
\hline
\multicolumn{2}{c|}{\textbf{Indoor Environment}}       & \multicolumn{1}{c|}{\textbf{}} & \multicolumn{2}{c}{\textbf{Outdoor Environment}}       \\ \cline{1-2} \cline{4-5} 
\multicolumn{1}{c}{\textbf{Models}} & \textbf{MDE (m)} & \multicolumn{1}{c|}{\textbf{}} & \multicolumn{1}{c}{\textbf{Models}} & \textbf{MDE (m)} \\ \cline{1-2} \cline{4-5} 
Baseline\cite{b14}                  & 7.90          &  & Baseline\cite{b15}                  & 398.40          \\
HADNN\cite{b18}                     & 14.93         &  & NN\cite{b17}                     & 357             \\
EA-CNN\cite{b19}                    & 8.34          &  & Ex. Trees\cite{b17}                 & 379             \\
\textbf{Proposed encoder} & \textbf{6.65} &  &  Proposed encoder & 361.21 \\
Proposed U-MLP   & 9.61 &  & \textbf{Proposed U-MLP}   & \textbf{341.94} \\ \hline
\end{tabular}
}
\end{table}

\section{CONCLUSION}\label{conclusion}

This paper presents an encoder-based TL and U-MLP model for accurate IoT localization using RSSI fingerprinting-based techniques. The proposed encoder-TL framework improves the baseline model by approximately 17.18\% in indoor environment and 9.79\% in outdoor environment, achieving MDEs of 6.65m and 361.21m, respectively. Similarly, our U-MLP achieved MDEs of 9.61m and 341.94m in indoor and outdoor environments, respectively. In the future, we plan to expand our research beyond these two datasets and explore a wider range of data sources and environments. Furthermore, with the increasing prevalence of IoT devices, our objective will be to further improve the security of our model to safeguard against potential threats and ensure uninterrupted performance.

\balance
\end{document}